\documentclass{jpsj2}

\def\runauthor{Online-Journal Subcommittee}

\title{%
Pair-Hopping Mechanism for Layered Superconductors
}

\author{%
Koichi Kusakabe 
}

\inst{%
Graduate School of Engineering Science, Osaka University,
Toyonaka, Osaka 560-8531, Japan
}

\recdate{\today}

\abst{%
We propose a possible charge fluctuation effect expected in 
layered superconducting materials. 
In the multireference density functional theory, 
relevant fluctuation channels for the Josephson coupling 
between superconducting layers include the interlayer pair hopping 
derived from the Coulomb repulsion. 
When interlayer single-electron tunneling processes are 
irrelevant in the Kohn-Sham electronic band structure calculation, 
the two-body effective interactions stabilize a superconducting phase. 
This state is also regarded as a valence-bond solid in a bulk electronic state. 
The hidden order parameters coexist with the superconducting order parameter 
when the charging effect of a layer is comparable to the pair hopping. 
Relevant material structures favorable for the pair-hopping mechanism 
are discussed. 
}

\kword{%
superconductivity, iron pnictide, cuprate, density functional theory, 
Haldane gap
}

\begin{document}
\maketitle

\section{Introduction}
The discovery of iron-pnictide superconductors gave us 
an interesting ground for testing theoretical approaches 
to analyze superconductivity. 
The first records of jump in the superconducting transition temperature 
in iron pnictides was observed in fluorine-doped 
LaFeAsO\cite{LaFeAsO,LaFeAsO2} when 
the discovery of fluorine-doped LaFePO\cite{LaFePO} 
fuelled a search for chemical trends in 
series of superconductors. 
Similar tests are also necessary to develop 
the theory of high-temperature superconductivity. 

In the present standard of theoretical approaches to analyze 
superconductivity in materials science, we may apply 
the strong-coupling theory of superconductivity 
starting from estimation of electron-phonon coupling constants 
{\it e.g.} by using 
the density-functional perturbation theory (DFPT).\cite{DFPT} 
The applications of DFPT 
in high-temperature element superconductors\cite{Tse,Uma} 
suggest that the technique enables us to obtain 
a reasonable estimation of the transition temperature $T_c$ 
when the essential pairing mechanism is 
the electron-phonon-interaction-mediated 
stabilization of superconductivity. 
The method may also predict how to enhance the $T_c$ 
of compound superconductors. 
As an example of its application, I and my coworkers have 
shown that a possible positive jump in $T_c$ in CaSi$_2$ is expected 
at a high pressure when a structural phase transition 
to the AlB$_2$ structure takes place.\cite{Nakanishi1} 

Recently, Jishi and Alahyaei have used this method to analyze the 
iron-based superconductors LiFeAs and NaFeAs.\cite{Jishi}
These 111 compounds may be ideal arsenides for 
the application of the strong-coupling theory, 
since clear superconducting transition 
temperatures of 18\,K for LiFeAs\cite{LiFeAs1,LiFeAs2} and 
9\,K for NaFeAs\cite{NaFeAs} 
are reported under ambient condition without doping. 
Another experimental report on NaFeAs suggests a higher transition temperature 
of above 12\,K without a clear indication of 
any coexisting magnetic ordering.\cite{LiFeAs-NaFeAs} 
Jishi and Alyahyaei unfortunately failed to obtain the transition temperature 
observed in experiments and concluded that the theory indicates a $T_c$ 
far below 1\,K owing to insufficient electron-phonon coupling constants. 

If one looks at a list of related materials,\cite{Ivanovskii} 
he or she might be motivated to perform another theoretical test 
in order to understand a chemical trend. 
The technique was pplied 
to NaFeAs by A. Nakanishi and 
he considered a chemical trend by studying 
other two supposed material structures: NaCoAs and NaNiAs. 
Using the calculation results, Kusakabe and Nakanishi 
studied the electronic structures of these compounds to derive 
a chemical trend.\cite{Nakanishi2} 
A key to understand the trend 
could be found in the direction to substitute Fe 
with other transition-metal elements, 
since electronic bands around the Fermi 
level are known to originate from iron 3d orbitals in Fe$_2$As$_2$ layers. 
However, note that 
there are only a few examples of nickel arsenides in the 1111 structure showing 
a finite $T_c$ of approximately 3\,K.\cite{Watanabe,Li} 
Fortunately, Nakanishi succeeded in observing the theoretical stability of 
the hypothetical compounds NaCoAs and NaNiAs, as well as NaFeAs, 
using first-principles structural optimization techniques\cite{Espresso} 
with the generalized gradient approximation (GGA).\cite{PBE} 
He applied the strong-coupling theory and estimated the transition 
temperature within a standard approach. 
By means of first-principles lattice dynamics, 
the superconducting transition temperature was estimated to be 
approximately 0.034\,K for NaFeAs, 
assuming that the system remained nonmagnetic. 
An interesting finding was the increase in  $T_c$ in the order of 
NaFeAs (0.034\,K), NaCoAs (0.127\,K), and NaNiAs (3.573\,K), 
although there might remain systematic errors due to the limitation in 
the computation. 
However, this possible chemical trend is in 
inverse relation to the experimental finding 
of $T_c=$9\,K or $T_c>$12\,K for NaFeAs and 
the $T_c$ for Co and Ni compounds 
not higher than that for this iron arsenide. 

As for the Kohn-Sham band structure on 
NaFeAs,\cite{Nakanishi2} the essential features around the Fermi level 
are rather similar to those of LiFeAs.\cite{Singh,Nekrasov} 
The effective bands in a nonmagnetic solution 
are almost dispersionless along the $c$ direction 
at approximately the Fermi level. 
Since electron and hole branches coexist and 
since DOS is reduced at approximately the Fermi level, we can conclude that 
the band structure has a semimetallic nature. That is, it shows 
strong two-dimensionality. 
Differently from LiFeAs, 
NaFeAs has two hole pockets. 
LiFeAs has three hole pockets 
on the two-dimensional plane including $\Gamma$-X-M lines in the 
first Brillouin zone. 
In addition, the two-dimensionality of the Fermi surface is much 
better for NaFeAs, and all of the pockets of NaFeAs except for one hole pocket 
around the $\Gamma$ point are rodlike.\cite{Nakanishi2} 

A crude realization of the rigid band picture 
among three band structures for NaMAs (M=Fe, Co, Ni) has been reported.
The correspondence of the major branches is clearly seen. 
Thus, the band structures of NaCoAs and NaNiAs are 
approximately a result of heavy-electron doping in the 
band structure of NaFeAs. 
In many iron arsenides, we find a semimetallic 
two-dimensional band structure.\cite{Singh-Du,Mazin,Mazin2} 
Following this interpretation, we can understand the reason why 
the electronic density of states (DOS)
at the Fermi energy increases 
when one considers the DOS's of NaCoAs and NaNiAs in comparison with 
that of NaFeAs. 
The Nakanishi data suggesting the increase 
in $T_c$ for NaCoAs and NaNiAs 
comes from the larger DOS together with the presence of 
more three-dimensional 
Fermi surfaces for these two compounds than for NaFeAs 
and the assumption of the electron-phonon-interaction-mediated 
superconductivity. 

To go one step further, 
we can start analyzing the 
characteristic topology of Fermi surfaces. 
The chemical trend of the Kohn-Sham band structure 
in the series of NaFeAs, NaCoAs, and NaNiAs in $P4/nmn$ suggests that 
rodlike two-dimensional Fermi pockets appear only in NaFeAs, 
while more three-dimensional characters for Co and Ni compounds 
indeed enhance $T_c$ if the strong-coupling theory with 
the electron-phonon coupling is assumed to be applicable. 
This result as well as the theoretical data on 
the above known first-principles electronic structure calculations of 
111 compounds leads us to a conclusion along the following line. 
An important ingredient tractable 
in the density functional theory (DFT) 
is charge fluctuation modes. 
If one of the modes becomes relevant on a two-dimensional Fermi surface, 
and if the fluctuation effect enhances the stability of a superconducting 
state, the theoretical approach starting from 
the standard Kohn-Sham scheme\cite{Kohn-Sham} would be feasible. 

In this study, we investigate 
charge fluctuation effects tractable in the multireference density 
functional theory (MR-DFT).\cite{Kusakabe-JPSJ,Kusakabe-JPCM} 
The two-dimensional electronic structures of iron arsenides 
found in Kohn-Sham band structure calculations 
actually suggest a Cooper-pair hopping mechanism in layered materials. 
We derive an effective theory of superconductivity. 
Important point is weak interlayer single-electron 
hopping processes in iron arsenides. 
In our MR-DFT formalism, charge fluctuation modes are 
introduced by fixing the Kohn-Sham single-particle description 
as a mean-field limit of the theory. 
Here, a simple effective Bosonic Hamiltonian is 
derived for layered superconductors. 
The model tells us that the 
formation of the valence-bond-solid state in a bulk superconducting 
state enhances the stability of the ordered state via the appearance of 
hidden order parameters. 
A hypothesis on the stabilized superconducting state will be addressed, 
where minimized charge fluctuation leading to the $S=1$ effective 
one-dimensional Heisenberg spin Hamiltonian 
is required for the most stable superconducting state in 
layered materials compared with other Heisenberg models with $S>1$. 
This picture is confirmed if we assume that quasi-particles 
in the layered superconductor highly correlate. 
Discretized quasi-particle spectrum expected in correlated electron systems 
ensures the effective spin Hamiltonian. 

To start discussion about the first-principles simulation method 
concrete, we address a new theory for 
the correlated electron systems called 
the density functional variational theory (DFVT). 
This is a simple variational method that always refers to a 
self-consistent solution given by MR-DFT. 
Finally, a means of applying the analysis techniques of DFVT 
to iron arsenide, as well as to other layered 
high-temperature superconductors including cuprates and 
MgB$_2$, will be addressed. 

\section{The Multireference Density Functional Theory}
In the standard DFT, 
the explicit form of the charge fluctuation is given by 
the energy density functional,\cite{Parr-Yang} 
\begin{eqnarray}
E_{fluc}[n]
&=&
\frac{e^2}{2}\int_0^1 d\lambda 
\int d^3r d^3r' 
\frac{1}{|{\bf r}-{\bf r}'|}
\langle \Psi_{n}^\lambda | 
(\hat{n}({\bf r})-n({\bf r}))(\hat{n}({\bf r}')-n({\bf r}'))
|\Psi_{n}^\lambda\rangle
\nonumber \\
&-&\frac{e^2}{2}\int_0^1 d\lambda \int d^3r d^3r' 
\frac{1}{|{\bf r}-{\bf r}'|}
\langle \Psi_{n}^\lambda | 
\sum_\sigma \psi^\dagger_\sigma({\bf r})
\psi_\sigma({\bf r}')\delta({\bf r}-{\bf r}')|\Psi_{n}^\lambda\rangle.
\label{fluc}
\end{eqnarray}
Here, $n({\bf r})$ is the electron density, 
$\psi_\sigma({\bf r})$ is the electron field operator 
for the spin $\sigma$, $\hat{n}({\bf r})=\psi^\dagger_\sigma({\bf r})
\psi_\sigma({\bf r})$, 
and $\Psi_{n}^\lambda$ is a minimizing wavefunction $\Psi'$ of 
the following reduced energy density functional 
\begin{equation}
F_\lambda[n]=\min_{\Psi'\rightarrow n}
\langle \Psi'|\hat{T}+\lambda \hat{V}_{ee}|\Psi' \rangle. 
\end{equation}
In the above definition, the kinetic energy operator $\hat{T}$ is 
\begin{equation}
\hat{T}=-\frac{\hbar^2}{2m} \int \! d^3r\, \sum_{\sigma} 
\lim_{{\bf r}' \rightarrow {\bf r}} 
\psi^\dagger_{\sigma}({\bf r}') \Delta_{\bf r} \psi_{\sigma}({\bf r}) \; ,
\end{equation}
where $m$ is the electron mass, 
and the Coulomb interaction is given in operator form as 
\begin{eqnarray}
\hat{V}_{\rm ee} &=& \frac{1}{2} \int \! d^3r \, d^3r' \,
\frac{e^2}{|{\bf r}-{\bf r}'|} \sum_{\sigma,\sigma'}
\psi^\dagger_{\sigma}({\bf r}) \psi^\dagger_{\sigma'}({\bf r}') 
\psi_{\sigma'}({\bf r}') \psi_{\sigma}({\bf r}). 
\end{eqnarray}
Although eq.\,(\ref{fluc}) is an exact expression, 
it is not easy to obtain insight into relevant charge fluctuation 
modes in a solid only by considering the use of 
this functional form. This is partly because the form is 
written in double integrals with respect to electron positions 
${\bf r}$ and ${\bf r}'$. However, we can see that 
the charge fluctuation may occur everywhere in the electron system. 
Once a crucial scattering due to the Coulomb fluctuation occurs 
between Fe$_{2}$As$_{2}$ layers, 
a pair scattering from one layer to another 
becomes a forward scattering. 

In our multireference density functional 
theory,\cite{Kusakabe-JPSJ,Kusakabe-JPCM} 
we can introduce part of the fluctuation explicitly by 
introducing a fluctuation term, 
$\langle \Psi | \hat{V}_{X_i} | \Psi \rangle$, with 
the self-interaction correction in the form 
\begin{eqnarray}
\label{X_i-term}
\langle \Psi | \hat{V}_{X_i} | \Psi \rangle
&=&
\sum_{n} X_i^{(n)} \langle \Psi |
\left\{
:\left( \hat{Y}_i^{(n)}-\langle \hat{Y}_i^{(n)}\rangle\right)^\dagger
\left( \hat{Y}_i^{(n)}-\langle \hat{Y}_i^{(n)}\rangle\right) :
\right.\nonumber \\
&&-\left.
:\left( \hat{Z}_i^{(n)}-\langle \hat{Z}_i^{(n)}\rangle\right)^\dagger
\left( \hat{Z}_i^{(n)}-\langle \hat{Z}_i^{(n)}\rangle\right):
\right\}
| \Psi \rangle . 
\end{eqnarray}
Since we may have a series of models, we use the notation 
$\langle \Psi | \hat{V}_{X_i} | \Psi \rangle$ with 
index $i$ specifying the number of the model. 
The notation $:\hat{O}:$  with an operator $\hat{O}$ denotes 
the normal ordering with respect to the creation and annihilation operators. 
The operators $\hat{Y}_i^{(n)}$ and $\hat{Z}_i^{(n)}$ may be given 
(i) by an expansion formula of the Coulomb operator 
around a fixed center and 
(ii) by a creation method for the Dirac character for the crystal. 
Definitions of each operators are 
given in the literature. 
Following the work,\cite{Kusakabe-Coulomb} 
we introduce the notation 
$\Phi_{pLM}({\bf r})=g_{pL}(r)Y_{LM}(\theta, \phi)$ and 
$\bar{\Phi}_{pLM}({\bf r})=\bar{g}_{pL}(r)Y_{LM}(\theta, \phi)$ 
using the spherical harmonics $Y_{LM}$ and 
a complete set of $g_{pL}(r)$ expanding the radial waves. 
Another function $\bar{g}_{pL}(r)$ is 
\begin{equation}
\bar{g}_{pL}(r)=\frac{1}{r^{L+1}}\int_0^rdr'(r')^L g_{pL}(r').
\end{equation}
In our Coulomb operator expansion formula, 
the $Y$ and $Z$ operators are given by, 
\begin{eqnarray}
\label{Yplm}
\hat{Y}_{pLM} 
&=& \frac{1}{\sqrt{2}}
\int d^3r \sum_\sigma 
\psi^\dagger_\sigma({\bf r})
\left(\Phi_{pLM}({\bf r})+\bar{\Phi}_{pLM}({\bf r})\right)
\psi_\sigma({\bf r})\;,\\
\label{Zplm}
\hat{Z}_{pLM} 
&=& \frac{1}{\sqrt{2}}
\int d^3r \sum_\sigma 
\psi^\dagger_\sigma({\bf r})
\left(\Phi_{pLM}({\bf r})-\bar{\Phi}_{pLM}({\bf r})\right)
\psi_\sigma({\bf r})\;.
\end{eqnarray}
The $n$th operators $\hat{Y}_i^{(n)}$ and $\hat{Z}_i^{(n)}$ 
in the $i$th model 
may be given by identifying a parameter set $p$, $L$, and $M$ 
as $n=(p,L,M)$. 
We have other possible expansions using a screened form of 
the Coulomb operator if we utilize DFVT given in the next section. 
The definitions of eqs.\,(\ref{Yplm}) and (\ref{Zplm}) 
are independent of the Kohn-Sham orbitals, 
which are used to expand the wavefunctions and the field operators 
in the creation and annihilation operators. 
This point gives an advantage to our formalism, 
because the scattering channels are defined before obtaining 
expressions of the Kohn-Sham orbitals. 

We also use the notation 
$n_\Psi$ to represent a density associated with a 
state $| \Psi\rangle$ as 
$n_\Psi({\bf r})\equiv \langle \Psi | \hat{n}({\bf r}) | \Psi \rangle$. 
The energy functional of the new extended Kohn-Sham scheme is 
\begin{eqnarray}
\label{ex-Kohn-Sham}
G_{X_i}[\Psi]&=&
\langle \Psi | \hat{T}+\hat{V}_{X_i} | \Psi \rangle
-\min_{\Psi'\rightarrow n_\Psi({\bf r})}
\langle \Psi' | \hat{T}+\hat{V}_{X_i} | \Psi' \rangle \nonumber \\
&+&F[n_\Psi]
+\int d^3r v_{\rm ext}({\bf r}) n_\Psi ({\bf r}) \nonumber \\
&=&
\langle \Psi | \hat{T}+\hat{V}_{X_i} | \Psi \rangle
+\frac{1}{2}\int\frac{n_\Psi({\bf r})n_\Psi({\bf r}')}
{|{\bf r}-{\bf r}'|}d^3rd^3r' 
\nonumber \\
&&+E_{\rm rxc}[n_\Psi] 
+\int d^3r v_{\rm ext}({\bf r}) n_\Psi ({\bf r}) \; . 
\end{eqnarray}
Here, we refer to 
the universal energy functional $F[n]$ given by 
\begin{eqnarray}
F[n]&=&\min_{\Psi'\rightarrow n}
\langle \Psi' | \hat{T}+\hat{V}_{\rm ee} | \Psi' \rangle \, .
\end{eqnarray}
The definition of $E_{\rm rxc}[n]$ is given 
by eq. (\ref{ex-Kohn-Sham}) itself. 
The new extended Kohn-Sham model 
is actually an effective many-body system. 

When we let some of $X_i^{(n)}$ be finite, if we replace 
$E_{\rm rxc}[n]$ with the GGA energy functional $E_{GGA}[n_\Psi]$, 
the model becomes a correlated Fermion model, which is 
defined by the approximated energy functional 
\begin{eqnarray}
\tilde{G}_{X_i}[\Psi]
&=&
\langle \Psi | \hat{T}+\hat{V}_{X_i}|\Psi \rangle 
+\frac{e^2}{2}
\int d^3r d^3r' \frac{n_\Psi({\bf r})n_\Psi({\bf r}')}{|{\bf r}-{\bf r}'|}
\nonumber \\
&&+E_{GGA}[n_\Psi]+
\int d^3r v_{ext}({\bf r})n_\Psi({\bf r}). 
\end{eqnarray}
The two-body scattering process happening in the $i$-th model 
is derived from the fluctuation term, 
which is divided into the effective two-body interaction and 
a counter term as 
\begin{eqnarray}
\frac{\delta \langle \Psi |\hat{V}_{X_i}|\Psi\rangle}{\delta \langle \Psi |}
&=&
{\cal H}_i^2 +  {\cal H}_{i,{\rm counter}}^1 \, . \\
{\cal H}^2_i
&=&
\sum_n X_i^{(n)}\left\{
:\left(\hat{Y}_{i}^{(n)}\right)^\dagger
       \hat{Y}_{i}^{(n)}:
-
:\left(\hat{Z}_{i}^{(n)}\right)^\dagger
       \hat{Z}_{i}^{(n)}:
\right\},
\\
{\cal H}_{i,{\rm counter}}^1
&=&
-
\sum_n X_i^{(n)}\left\{
\langle 
\left(\hat{Y}_i^{(n)}\right)^\dagger
\rangle 
\cdot 
      \hat{Y}_i^{(n)}
+
\left(\hat{Y}_i^{(n)}\right)^\dagger
\cdot 
\langle 
      \hat{Y}_i^{(n)}
\rangle 
\right.\nonumber \\ 
&&\left.
-
\langle 
\left(\hat{Z}_i^{(n)}\right)^\dagger
\rangle 
\cdot 
      \hat{Z}_i^{(n)}
-
\left(\hat{Z}_i^{(n)}\right)^\dagger
\cdot 
\langle 
      \hat{Z}_i^{(n)}
\rangle \right\}. 
\label{counter-model-many-body-EKS}
\end{eqnarray}
Since we have the exchange-correlation potential for GGA as 
\begin{equation}
\mu_{GGA}({\bf r})
=\frac{\delta E_{GGA}[n]}{\delta n({\bf r})},
\end{equation}
a secular equation is derived by imposing the normalization 
condition of $|\Psi\rangle$ using the Lagrange multiplier $E$ as, 
\begin{eqnarray}
\frac{\delta \tilde{G}_{X_i}[\Psi]}{\delta \langle \Psi |}
&=&{\cal H}^{\rm eff}_i |\Psi\rangle = E|\Psi\rangle, \\
\label{model-many-body-EKS}
{\cal H}^{\rm eff}_i &=& {\cal H}^1 + {\cal H} ^2_i
 + {\cal H}_{i,{\rm counter}}^1\; ,\\
{\cal H}^1 &=& \hat{T}
+\int \bar{v}_{\rm eff}({\bf r})\hat{n}({\bf r})d^3r \; . 
\end{eqnarray}
The effective single particle potential 
$\bar{v}_{\rm eff}({\bf r})$ is given by 
\begin{equation}
\label{model-Effective-potential}
\bar{v}_{\rm eff}({\bf r}) = 
\int \frac{n({\bf r}')}{|{\bf r}-{\bf r}'|}d^3r' 
+ \mu_{GGA}({\bf r})
+ v_{\rm ext}({\bf r}) \; .
\end{equation}
We have a potential problem given as, 
\begin{equation}
\label{model-one-body-EKS}
\left\{-\frac{\hbar^2}{2m}\Delta_{\bf r} +\bar{v}_{\rm eff}({\bf r})\right\}
\chi_{m,{\bf k},k_z}({\bf r})
=\varepsilon_{m,{\bf k},k_z}\chi_{m,{\bf k},k_z}({\bf r}),
\end{equation}
in which Bloch orbitals $\chi_{m,{\bf k},k_z}({\bf r})$ 
are determined to be normalized and orthonormal in a crystal phase. 
Here, the Kohn-Sham orbital 
is specified by a two-dimensional wave vector, ${\bf k}$, 
another wave vector, $k_z$, along the $c$ axis, and the band index $m$. 

We do not explicitly write the $i$ dependence on 
${\cal H}^1$. However, it is implicitly dependent on 
the fluctuation term 
$\langle \Psi | \hat{V}_{X_i} | \Psi \rangle$ through 
the self-consistency on the charge density 
$n({\bf r})=n_\Psi({\bf r})$. 
We may have another definition of a single-particle problem 
by including the nonlocal potential part 
and/or mean-field part coming from 
$\langle \hat{Y}_i^{(n)} \rangle$ 
and $\langle \hat{Z}_i^{(n)} \rangle$, {\it i.e.} 
the 1-body counterterm ${\cal H}_{i,{\rm counter}}^1$. 
However, this shift only changes the definition of 
single-particle orbitals used to determine the many-body problem. 
The charge density $n_\Psi({\bf r})$ determines the expectation values 
$\langle \hat{Y}_i^{(n)}\rangle$ and $\langle \hat{Z}_i^{(n)}\rangle$. 
Thus, if $n_\Psi({\bf r})$ is almost unchanged in a self-consistent loop, 
these values also remain unchanged. 
Even if there are many-body correlation effects 
creating a few meV of gap in the model, 
the essential features of the GGA band structure are 
unaffected by only the part ${\cal H}_{i,{\rm counter}}^1$. 
An essential change may occur via the correlation effects 
appearing as a slight change in $n_\Psi({\bf r})$ 
and a large change in variational energy 
via ${\cal H} ^2_i$. 

When $X_i^{(n)}=0$ for all $n$ values, 
we obtain a secular equation of the Kohn-Sham equation in GGA 
given by eq.\,(\ref{model-one-body-EKS}). 
The ground state of this model is slightly shifted from the final state, 
when we consider a correlated electron system. 
The obtained single-particle description defines 
the Kohn-Sham band structure. Introducing a proper Fourier transformation 
for a selected set of bands, we immediately obtain the Wannier representation, 
allowing us to rewrite ${\cal H}^{\rm eff}_i$ in 
a second-quantized form of a tight-binding model. 

In the GGA calculation, we may consider the spin density by 
introducing a spin-dependent GGA functional. 
If we adopt the spin-GGA scheme, 
the Kohn-Sham equation becomes spin-dependent 
for a magnetic solution. 
For the formulation of MR-DFT, however, spin-independent Kohn-Sham orbitals 
are very useful for the discussion of magnetism and superconductivity. 
This is because the correlation effects are described by 
the multireference variational states, 
{\it i.e.} the multi-Slater determinants, in MR-DFT. 
Thus our starting point is a paramagnetic state obtained by 
a nonmagnetic GGA calculation. 

\section{Pair-Hopping Mechanism}
In the first-principles study of compound superconductors, 
we may apply the Kohn-Sham scheme\cite{Kohn-Sham} 
of the density functional theory (DFT)\cite{Hohenberg-Kohn} 
as a starting point. 
Thus, we investigate Kohn-Sham band structures, which 
are known in the literature or found in actual calculation results. 
The important points considered here are the structures of the Fermi surface 
and the dimensionality of the low-energy branches of the band structure. 

The pictures drawn from the study summarized in 
the last section are as follows. 
We have several examples of iron arsenides, 
which have two-dimensional Fermi surfaces, 
within the Kohn-Sham scheme. 
Details of the Fermi surfaces actually depend on the type of material. 
However, in general, 
the GGA band structure suggests a picture of a stack of Fe$_2$As$_2$ layers 
weakly coupled by single-particle tunneling at the Fermi level. 

If we expect Fermi instability due to the remaining Coulomb fluctuation 
modes lost in a mean-field approach, 
or if relevant fluctuation only shifts the electronic structure 
around the Fermi level by opening a gap of a few tens meV, 
only a slight change in the total charge density due to 
appearance of secondary order parameters is expected. 
Then, a picture of the layered two-dimensional Fermi gas system 
should remain as a starting one-body mean-field state 
even in the final many-body solution affected by correlation effects. 

In this picture, we need to incorporate effective two-body interactions 
both within a layer and between layers. 
Here, note that there is no interband single-particle 
hopping process, since the single-particle part is diagonal 
in the band index. 
A step to maintain self-consistency is necessary after solving 
the many-body problem given by the effective two-body repulsive interactions, 
since the charge redistribution might or might not modify the effective 
single-particle excitation spectrum at approximately the Fermi level. 

The stacking of two-dimensional systems in one direction 
forms a layered bulk system. 
The system becomes a correlated electron system, 
because the single-electron tunneling process is reduced 
owing to the two-dimensionality and because 
possible charge fluctuations in a layer should be suppressed 
also owing to the localized nature of Fe 3d orbitals in the superconductor. 
This picture is tractable by constructing a one-dimensional Wannier 
representation from the Kohn-Sham orbitals. 
The band dispersions around the Fermi level are almost flat along 
the $c$-direction, so that the one-dimensional Wannier representation 
is natural. 

The motion of electrons in a layer is described by 
a two-dimensional electron gas model. 
Here, a two-dimensional wave vector ${\bf k}$ and a 
band index $m$, or a combined index $j=(m,{\bf k})$, 
are used to specify each Wannier state in the $l$-th layer. 
In the two-dimensional system 
with multicolored Fermion quasi-particles in the layer, 
a reduced model might be well-described by a known 
two-dimensional model.\cite{Mazin,Kuroki} 
In such a model, the effective electron-electron interaction parameters 
would specify intralayer scattering processes. 
A superconducting fluctuation may occur, if we consider 
the two-body scattering processes due to charge or spin fluctuations, 
which is not explicitly counted in the GGA calculation. 
The electron-phonon interaction may contribute to the stability of 
the superconducting fluctuation in the layer. 
For our next discussion, however, intralayer effective attractions 
can originate from any mechanism, as far as it 
effectively supports the formation of 
the precursors of the bulk superconducting 
order parameter. 

Here, the important driving force derived in this study 
is the interlayer pair-hopping processes between the layers. 
When the Kohn-Sham orbitals are determined using 
the effective exchange-correlation potential, $v_{xc}({\bf r})$, 
these charge fluctuation modes are not included explicitly. 
In MR-DFT, the modes are explicitly introduced using two-body operators. 
The two-body processes are given by the repulsive nature of 
the Coulomb electron-electron scattering. 
If we have any effective attractive interaction at a position 
between the layers, the mechanism is lost. 
The origin of the charge fluctuation effect 
will now be discussed in detail. 

We consider the Wannier representation of orbitals localized in a layer. 
The orbital wavefunction has a representation 
$\phi_{jl}({\bf r})$ without an explicit spin dependence. 
Actually, we can use a proper unitary transformation, 
{\it e.g.}, the Wannier transformation, to create 
$\phi_{jl}({\bf r})$ from $\chi_{m,{\bf k},k_z}({\bf r})$. 
Here, $l$ denotes an index of a layer and 
$j$ represents a set of indexes ($m,{\bf k}$). 
By associating creation and annihilation operators, $c^\dagger_{jl}$ 
and $c_{jl}$ are defined by the Canonical anti-commutation relation, 
\begin{equation}
\{c^\dagger_{jl\sigma}, c_{j'l'\sigma'} \} 
= \delta_{jj'} \delta_{l,l'} \delta_{\sigma,\sigma'}. 
\end{equation}

As a relevant perturbation for the GGA band structure, we consider 
scattering processes coming from the charge fluctuation in 
$\langle \Psi | \hat{V}_{X_i} | \Psi \rangle$. 
In the expression of the $Y$ and $Z$ operators, 
we have a pair of field operators, namely, 
$\psi^\dagger_\sigma({\bf r})$ and $\psi_\sigma({\bf r})$. 
These operators are expanded in the localized orbitals 
$\phi_{jl}({\bf r})$ and $c^\dagger_{jl\sigma}$, $c_{jl\sigma}$. 
Thus, we notice that 
we have a double summation in the definitions of 
$\hat{Y}_i^{(n)}$ and $\hat{Z}_i^{(n)}$.\cite{Kusakabe-Coulomb} 
One is for the conserved quantities and the other is with respect to 
the orbitals in the Wannier form. 
Here, the center for the Coulomb expansion formula is not 
necessarily identical to a Wannier center. 
Depending on the symmetry of an electron pair 
in both the initial and final states, the 
proper selection of $n=(p, L, M)$ is given to minimize the energy 
using DFVT. We can then identify relevant fluctuation terms, 
in which $\hat{Y}_i^{(n)}$ may connect a localized orbital 
in the low-energy bands at the Fermi energy with 
other semilocalized orbitals, leading to 
the pair hopping between Wannier centers. 

When a superconducting order parameter 
$\bar{\Delta}_l({\bf r},{\bf r}')\equiv 
\sum_{\langle jj'\rangle} 
\phi_{jl}({\bf r}) \phi_{j'l}({\bf r}')
\langle c_{jl\uparrow} c_{j'l\downarrow} \rangle$ 
is expected to be finite, 
each term in $\langle \Psi|\hat{V}_{X_i}|\Psi\rangle$ 
can be re-expressed as below. 
As an example, we consider the expression 
\begin{eqnarray}
\lefteqn{
X_i^{(n)}
\langle \Psi |
:\left( \hat{Y}_i^{(n)}-\langle \hat{Y}_i^{(n)}\rangle\right)^\dagger
\left( \hat{Y}_i^{(n)}-\langle \hat{Y}_i^{(n)}\rangle\right) :
| \Psi \rangle
}
\nonumber \\
&=&
X_i^{(n)}\left\{
\langle \Psi |
:\left( \hat{Y}_i^{(n)}\right)^\dagger
\hat{Y}_i^{(n)}:
|\Psi \rangle
-
\langle \left( \hat{Y}_i^{(n)}\right)^\dagger \rangle 
\langle \hat{Y}_i^{(n)}\rangle 
\right\}
\nonumber \\
&=&
\sum_{l_1l_2l_3l_4,j_1j_2j_3j_4}
\sum_{\sigma,\sigma'}
\frac{X_i}{2}
\int d^3r \int d^3r' 
\phi^*_{j_1l_1}({\bf r})
\left(\Phi_{pLM}({\bf r})+\bar{\Phi}_{pLM}({\bf r})\right)^*
\phi_{j_4l_4}({\bf r})
\nonumber \\
&\times&
\phi^*_{j_2l_2}({\bf r})
\left(\Phi_{pLM}({\bf r})+\bar{\Phi}_{pLM}({\bf r})\right)
\phi_{j_3l_3}({\bf r})
\langle:
c^\dagger_{j_1l_1\sigma}
c_{j_4l_4\sigma}
c^\dagger_{j_2l_2\sigma'}
c_{j_3l_3\sigma'}
:\rangle
-
X_i^{(n)}
\langle \left( \hat{Y}_i^{(n)}\right)^\dagger \rangle 
\langle \hat{Y}_i^{(n)}\rangle 
\nonumber \\
&=&
\sum_{l_1=l_2=l_3=l_4=l,j_1j_2j_3j_4}
\sum_{\sigma,\sigma'}
g_{n;llll,j_1j_2j_3j_4}
\langle :c^\dagger_{j_1l\sigma}
c_{j_4l\sigma}
c^\dagger_{j_2l\sigma'}
c_{j_3l\sigma'}: \rangle
\nonumber \\
&+&
\sum_{l_1=l_2=l\neq l_3=l_4=l',j_1j_2j_3j_4}
\sum_{\sigma,\sigma'}
g_{n;lll'l',j_1j_2j_3j_4}
\left\{
\langle c^\dagger_{j_1l\sigma}
c^\dagger_{j_2l\sigma'}\rangle
\langle c_{j_3l'\sigma'}
c_{j_4l'\sigma}\rangle
\right.
\nonumber \\
&+&
\left.
\langle \left(
c^\dagger_{j_1l\sigma}
c^\dagger_{j_2l\sigma'}
-
\langle c^\dagger_{j_1l\sigma}
c^\dagger_{j_2l\sigma'}\rangle
\right)
\left(
c_{j_3l'\sigma'}
c_{j_4l'\sigma}
-
\langle c_{j_3l'\sigma'}
c_{j_4l'\sigma}\rangle
\right)\rangle
\right\}
\nonumber \\
&+&
\sum_{l_1\neq l_2 \, {\rm or} \, l_3\neq l_4,j_1j_2j_3j_4}
\sum_{\sigma,\sigma'}
g_{n;l_1l_2l_3l_4,j_1j_2j_3j_4}
\langle:
c^\dagger_{j_1l_1\sigma}
c_{j_4l_4\sigma}
c^\dagger_{j_2l_2\sigma'}
c_{j_3l_3\sigma'}
:\rangle
\nonumber \\
&-&
X_i^{(n)}
\langle \left( \hat{Y}_i^{(n)}\right)^\dagger \rangle 
\langle \hat{Y}_i^{(n)}\rangle .
\label{Y-term}
\end{eqnarray}
Here, the coefficient $g_{n;l_1l_2l_3l_4,j_1j_2j_3j_4}$ is 
given by 
\begin{eqnarray}
g_{n;l_1l_2l_3l_4,j_1j_2j_3j_4}
&=&
\frac{X_i^{(n)}}{2}
\int d^3r \int d^3r' 
\phi^*_{j_1l_1}({\bf r})
\left(\Phi_{pLM}({\bf r})+\bar{\Phi}_{pLM}({\bf r})\right)
\phi_{j_4l_4}({\bf r})
\nonumber \\
&\times&
\phi^*_{j_2l_2}({\bf r}')
\left(\Phi_{pLM}({\bf r}')+\bar{\Phi}_{pLM}({\bf r}')\right)
\phi_{j_3l_3}({\bf r}')
\, .
\end{eqnarray}
A similar expansion is also given for terms with the $\hat{Z}_i^{(n)}$ 
operators. 
Thus, the effective two-body Hamiltonian 
${\cal H}_i^2$ is divided into three parts: 
(1) the intralayer two-body Hamiltonian ${\cal H}_{i,{\rm intra}}^2$ 
(with $g_{n;llll,j_1j_2j_3j_4}$), 
(2) the interlayer pair hopping 
Hamiltonian ${\cal H}_{i,{\rm inter-pair}}^2$
(with $g_{n;lll'l',j_1j_2j_3j_4}$), and 
(3) other 2-body terms ${\cal H}_{i,{\rm inter-res}}^2$. 

The first category ${\cal H}_{i,{\rm intra}}^2$ 
contains terms interpreted as the on-site Hubbard repulsion. 
However, to obtain an  explicit form, we need to introduce 
another Wannier transformation to have Wannier orbitals 
localized around a Wannier center in a layer.\cite{Marzari} 
This process might be difficult, since 
we need to introduce unitary transformation in a rather wide 
energy window over a few eV. 
In our discussion, this process is not required to derive 
an effective low-energy model of high-temperature superconductors. 
In ${\cal H}_{i,{\rm intra}}^2$, 
we also have intralayer exchange interaction, 
intralayer off-site repulsion, and intralayer pair hopping. 
These effective interactions may induce the spin fluctuation effect 
as well as the charge fluctuation effect. 
For a general discussion, we do not specify the detailed 
form of ${\cal H}_{i,{\rm intra}}^{2}$, thereby allowing 
a BCS model Hamiltonian for the electron-phonon mechanism, 
a correlated two-dimensional electron model, 
and a model considering both effects. 
In our pair-hopping mechanism, 
${\cal H}_{i,{\rm inter-pair}}^{2}$ plays a relevant role. 

Note that $X_i^{(n)}$ is finite 
for a selected set of $p$, $L$, and $M$ values. 
The second category ${\cal H}_{i,{\rm inter-pair}}^2$ is 
derived from a representation as 
a mean-field term plus the fluctuation term 
using the superconducting order as in eq.\,(\ref{Y-term}). 
The third category ${\cal H}_{i,{\rm inter-res}}^2$ 
contains interlayer exchange interaction, inter-layer 
correlated hopping terms and 
interlayer diagonal charge fluctuation.  
We express the interlayer two-body Hamiltonian 
as ${\cal H}_{i,{\rm inter}}^2
={\cal H}_{i,{\rm inter-pair}}^2+{\cal H}_{i,{\rm inter-res}}^2$. 

The effective model may be written in a second quantized form as 
\begin{eqnarray}
{\cal H}_{\rm eff} &=& 
 {\cal H}_{\rm intra}^{1}+{\cal H}_{i,{\rm intra}}^{2}
+{\cal H}_{\rm inter}^{1}+{\cal H}_{i,{\rm inter}}^{2}
+{\cal H}_{i,{\rm counter}}^1, \\
{\cal H}_{\rm intra}^{1}&=&
\sum_{l} 
\sum_{\langle jj'\rangle} \sum_\sigma
t_{jj'}^{(l)} \left\{ c^\dagger_{jl\sigma} c_{j'l\sigma} + {\rm H.c.} \right\}, 
\\
{\cal H}_{\rm inter}^{1}&=&
\sum_{l\neq l'} \sum_{\langle jj'\rangle} \sum_\sigma
t_{jj'}^{(ll')} \left\{ c^\dagger_{jl\sigma} 
c_{j'l'\sigma} + {\rm H.c.} \right\}, 
\end{eqnarray}
where 
${\cal H}_{\rm intra}^{1}$ 
is the intralayer single-body Hamiltonian and 
${\cal H}_{\rm inter}^{1}$ 
is the interlayer single-body Hamiltonian. 
The index $l$ specifies an $l$-th layer and 
$j$ represents a $j$-th orbital in the $l$-th layer. 
Note again that the $j$-th orbital is 
a Wannier representation made by one-dimensional 
Frourier transformation from the Bloch waves. 
Thus, we can always say that 
the interlayer single-body process for 
the Wannier states around the Fermi level 
is negligible for iron-arsenide superconductors. 
At energy levels above the Fermi level, however, 
we also have finite bandwidths for the GGA band structure 
in the $c$ direction. 
This picture is very important for the discussion below. 

We now construct a standard model of 
high-temperature superconductivity. 
Two steps are necessary. First is the 
construction of a mean-field description and 
second is the derivation of the many-body effective Hamiltonian. 
We consider the effective intralayer Hamiltonians 
${\cal H}_{\rm intra}^{1}$ and ${\cal H}_{i,{\rm intra}}^{2}$, which 
induce superconducting fluctuation. 
However, owing to their explicit two-dimensionality, 
${\cal H}_{\rm intra}^{1}+{\cal H}_{i,{\rm intra}}^{2}$ cannot 
induce bulk superconductivity by itself. 
We then introduce the mean-field description of 
a two-dimensional superconducting state on a layer in 
a self-consistent field of other layers. 
Here, let us consider singlet superconductivity. 
The order parameter 
$\bar{\Delta}_l({\bf r},{\bf r}')\equiv 
\sum_{\langle jj'\rangle} 
\phi_{jl}({\bf r}) \phi_{j'l}({\bf r}')
\langle c_{jl\uparrow} c_{j'l\downarrow} \rangle$ 
can have a finite value around the $l$-th layer. 
One important point is that 
$\bar{\Delta}_l({\bf r},{\bf r}')$ may change its phase as $(-1)^l$, but 
the following theory also allows a constant phase factor for all layers. 
The final superconducting phase should be determined by 
a variational determination method. 

We may consider two-body fluctuation, which induces pair-hopping processes 
between neighboring layers. The process comes from 
${\cal H}_{i,{\rm inter-pair}}^{2}$. 
This term results in a pair field Hamiltonian for the $l$-th layer. 
\begin{eqnarray}
{\cal H}_{\rm pf}^{l}
&=&
\sum_n X_i^{(n)}
\int d^3r d^3r' 
\sum_{l'\neq l}
\sum_{jj'} 
\left\{ 
\bar{\Delta}_{l'}^*({\bf r},{\bf r}')
\left(\Phi_{pLM}({\bf r})^* \times \bar{\Phi}_{pLM}({\bf r}')
\right.
\right.
\nonumber \\
&&
\left.
\left.
     +\bar{\Phi}_{pLM}({\bf r})^* \times \Phi_{pLM}({\bf r}')\right)
\phi_{jl}({\bf r}) \phi_{j'l}({\bf r}')
c_{jl\uparrow} c_{j'l\downarrow}
+{\rm H.c.}
\right\}
\nonumber \\
&=&
\sum_{jj'} 
\left\{ 
\bar{\Delta}_{jj'l}^*
c_{jl\uparrow} c_{j'l\downarrow}
+{\rm H.c.}
\right\}
. \label{Pair-field1}
\end{eqnarray}
We have introduced an effective coupling constant $X_i^{(n)}$ 
for each scattering channel described by 
$:\left(\hat{Y}_{i}^{(n)}\right)^\dagger
      \hat{Y}_{i}^{(n)}
     -\left(\hat{Z}_{i}^{(n)}\right)^\dagger
      \hat{Z}_{i}^{(n)}:$. 
It can be derived from the Coulomb kernel 
$\displaystyle \frac{e^2}{2|{\bf r}-{\bf r}'|}$, however,  
in the multireference density functional theory, 
effective coupling can be optimized. 
By applying the fluctuation reference method, 
we should determine the parameter for reproducing another precise calculation. 
Or the effective Hamiltonian can be determined using DFVT, 
which will be addressed in the next section. 
A relevant point in our discussion is that 
the model is derived using the multireference density functional theory. 

The above derivation of eq.\,(\ref{Pair-field1}) is 
given by direct pair hopping. 
This Coulomb off-diagonal element is 
negligible for LaFeAsO$_{1-x}$F$_x$, 
since the neighboring two iron layers are widely separated by 
a La$_2$(O$_{1-x}$F$_{x}$)$_2$ layer. 
However, we have a pair-tunneling process across the insulating layer. 
We call it the super pair tunneling. 
To be precise, we show the construction step of the effective 
Hamiltonian for doped LaFeAsO. 
We can perform the non-magnetic GGA calculation of, {\rm e.g.}, 
LaFeAsO$_{0.875}$F$_{0.125}$ using a super cell with 
an optimized atomic position. 
The GGA band structure reveals the appearance of well-localized 
3d bands of iron at the Fermi level. 
Both electron and hole pockets are created from 
the localized 3d orbitals. 
These center bands have a clear two-dimensionality. 
Above these bands, at approximately 3 $\sim$ 4\,eV 
higher than the Fermi level, we have delocalized bands that consist of 
non-$s$ orbitals at La sites and void sites between a Fe$_2$As$_2$ layer 
and a La$_2$O$_2$ layer. 
These extended bands are formed by the hybridization between 
these high energy levels and iron 3d orbitals, 
so that localized 3d orbitals 
connect to higher levels using finite matrix elements 
by two-body Coulomb scattering processes. 
With the help of these orbitals, 
an indirect pair-hopping process from a Fe$_2$As$_2$ layer to 
the next layer by Coulomb off-diagonal elements is allowed. 
This second-order perturbation process is relevant. 
Its effective form finally becomes the same as 
${\cal H}_{i,{\rm inter-pair}}^{2}$, if we replace 
the interaction kernel with the effective one. 
We could also have a higher-order contribution from 
other terms in ${\cal H}_{i,{\rm inter-res}}^{2}$ 
for the pair-field Hamiltonian. 
However, it is important to note 
that ${\cal H}_{i,{\rm inter}}^{2}$ is always necessary to have 
the energy reduction in the superconducting state, 
since ${\cal H}_{\rm inter}^1$ is negligible at the Fermi energy 
and since ${\cal H}_{\rm inter}^1$ is diagonal in the band index. 

For 11 compounds, 
we may consider only orbitals in a Fe$_2$Se$_2$ 
layer or a Fe$_2$Te$_2$ layer. 
In these systems, a direct pair hopping from one layer to the next layer 
is possible via Coulomb repulsion. 
Thus, we have two different categories of the pair-hopping mechanism. 
The first is the direct pair hopping and the second is 
the indirect super pair tunneling. 
Both of the processes require a finite amplitude for 
the pair hopping from a localized orbital to another well-defined orbital. 

The inclusion of the pair field necessarily results in 
a Josephson coupled superconducting state as a variational ground state. 
Its local wavefunction is given by the following effective Hamiltonian 
for the $l$-th layer: 
\begin{equation}
{\cal H}_{\rm eff}^{l}
=
\sum_{\langle jj'\rangle} \sum_\sigma
t_{jj'}^{(l)} \left\{ c^\dagger_{jl\sigma} c_{j'l\sigma} + {\rm H.c.} \right\} 
+{\cal H}_{i,{\rm intra},l}^{2}
+{\cal H}_{\rm pf}^{l}, 
\end{equation}
where the intralayer two-body Hamiltonian
${\cal H}_{i,{\rm intra},l}^{2}$ for the $l$-th layer 
can be either electron-electron-interaction originated, 
electron-phonon-interaction originated, or their combination. 
The pair-hopping processes producing ${\cal H}_{\rm pf}^{l}$ 
can give energy gain to the Coulombic electron system, 
although a finite energy loss 
occurs when single-particle tunneling processes 
between layers induced by ${\cal H}_{\rm inter}^{1}$ are 
terminated to have a variational state, even if they are negligible. 

In the mean-field description, we are able to 
obtain a mean-field solution, which have two order parameters: 
$n({\bf r})$ and $\bar{\Delta}_l({\bf r},{\bf r}')$. 
In the MR-DFT formalism, 
appearance of $\bar{\Delta}_l({\bf r},{\bf r}')$ 
affects the single-particle momentum distribution and 
its Fourier transform, {\it i.e.} $n({\bf r})$. 
Therefore, we can interpret that $n({\bf r})$ becomes 
$\bar{\Delta}_l$-dependent. 
For the derivation, however, we need techniques for 
discretizing several continuous variables to have 
a tractable model in MR-DFT simulation. 
Here, we would rather move onto another effective theory 
to consider the superconducting phase derived from 
interlayer pair-hopping processes.

When a mean-field wavefunction of the layered material is 
obtained in the form 
\begin{equation}
|\Psi^{(0)}\rangle 
= \prod_{l}\prod_{m{\bf k}}
\otimes
\left(u_{m\bf k}^{(l)}+v_{m\bf k}^{(l)}b^\dagger_{m{\bf k}l}\right)
|0\rangle,
\label{Non-correlated-super-state}
\end{equation}
we can consider an explicit pair hopping. 
Here, $b^\dagger_{m{\bf k}l}=
c^\dagger_{m{\bf k}l\uparrow}
c^\dagger_{m-{\bf k}l\downarrow}$ with 
$c^\dagger_{m{\bf k}l\sigma}=c^\dagger_{jl\sigma}$, 
where $m$ is the band index and ${\bf k}$ is the two-dimensional wave vector. 
The real factor $u_{m\bf k}^{(l)}$ and another complex factor, 
$v_{m\bf k}^{(l)}$, satisfy 
$\left(u_{m\bf k}^{(l)}\right)^2+|v_{m\bf k}^{(l)}|^2=1$. 
However, note that, 
even if we have a correlated superconducting state 
$|\Psi^{(0)}\rangle$ with an expression other 
than eq.\,(\ref{Non-correlated-super-state}), 
we can always construct a Bogoliubov-Valatin transformation 
using the superconducting order parameter 
$\bar{\Delta}_l({\bf r},{\bf r}')$. 

\begin{figure}[tb]
\begin{center}
\includegraphics[width=10.0cm]{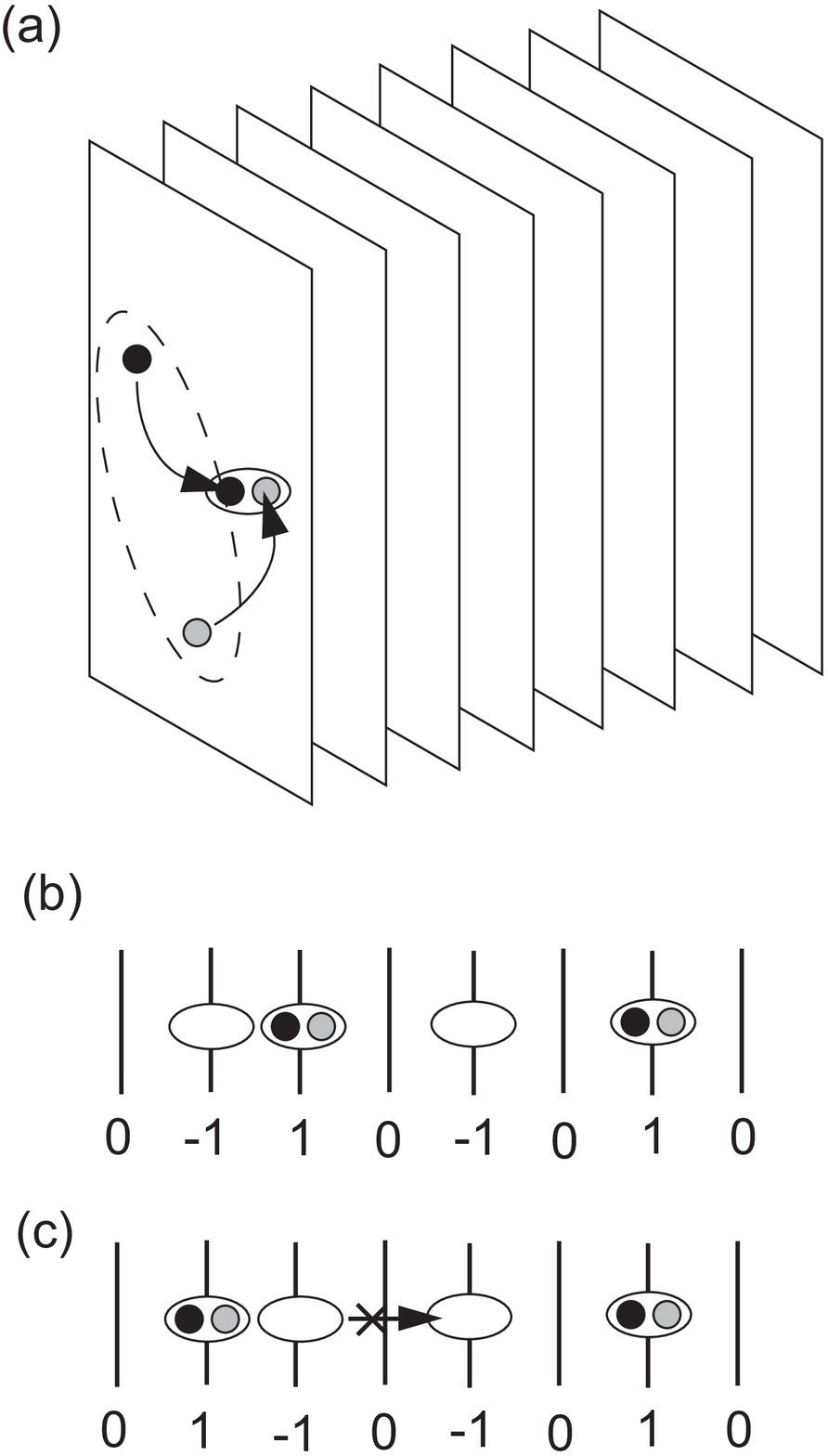}
\end{center}
\caption{
Minimal model of the high-temperature superconductivity 
in a layered material. 
(a) In the layered superconductor, 
a Cooper pair may hop via a two-body pair-hopping process 
into the next layer, but single-electron tunneling processes 
are suppressed. (b) Extra pairs and resulting vacancies 
may be created by the hopping process with charge fluctuation. 
(c) Neighboring pairs or neighboring vacancy layers are 
energetically forbidden. Thus, we have a hidden string 
order in the Haldane phase. 
}
\label{Fig-1}
\end{figure}

Now we go to the final step in order to take the fluctuation effect 
into account and to obtain the hidden order parameter and the energy gap. 
When a pair described by $b^\dagger_{m{\bf k}l}$ 
hops via the annihilation operation
$\Delta_l({\bf r},{\bf r}')\equiv 
\sum_{\langle jj'\rangle} 
\phi_{jl}({\bf r}) \phi_{j'l}({\bf r}')
c_{jl\uparrow} c_{j'l\downarrow}$, 
the layer loses two electrons and the next layer obtains 
these electrons (see Fig.\,\ref{Fig-1}\,(a)). 
Each layer should keep its charge neutrality, except for 
local charge fluctuation. 
Thus, if charge fluctuation effects are introduced into 
$|\Psi^{(0)}\rangle$, 
an effective action appears for this motion of 
pairs in the array of layers. 
(Fig.\,\ref{Fig-1}(b).) 
Then, we obtain a perturbed state 
$|\Psi^{(1)}\rangle$, which is determined by 
the effective action of the pairs. 
We derive the effective action using only a unitary transformation 
without referring to these supposed state vectors. 

The charging effect should be of the same order of magnitude 
as the pair-hopping process. 
If the intralayer coherence is well kept, 
but if the interlayer single-electron hopping processes are not relevant, 
the screening effect expected for the system mainly occurs in the layer. 
The condition for 
${\cal H}_{\rm inter}^{1}$ is consistent with this picture. 
Then, the number of pairs allowed to hop at one time 
in a fluctuation process is restricted to be very small. 

To justify this discussion, we introduce 
the Josephson-Bardeen modification\cite{Josephson,Bardeen} of 
the Bogoliubov-Valatin transformation: 
\begin{eqnarray}
\hat{\gamma}_{em{\bf k}l\uparrow}^\dagger
&=&
 u_{m{\bf k}}^{(l)} c_{m{\bf k}l\uparrow}^\dagger
-v_{m{\bf k}}^{(l)} S_l^* c_{m-{\bf k}l\downarrow},
\nonumber \\
\hat{\gamma}_{hm{\bf k}l\uparrow}^\dagger
&=&
 u_{m{\bf k}}^{(l)} S_l c_{m{\bf k}l\uparrow}^\dagger
-v_{m{\bf k}}^{(l)} c_{m-{\bf k}l\downarrow}
=S_l\hat{\gamma}_{em{\bf k}l\uparrow}^\dagger,
\nonumber \\
\hat{\gamma}_{em{\bf k}l\downarrow}^\dagger
&=&
 u_{m{\bf k}}^{(l)} c_{m-{\bf k}l\downarrow}^\dagger
+\left(v_{m{\bf k}}^{(l)}\right)^* S_l^* c_{m{\bf k}l\uparrow},
\nonumber \\
\hat{\gamma}_{hm{\bf k}l\downarrow}^\dagger
&=&
 u_{m{\bf k}}^{(l)} S_l c_{m-{\bf k}l\downarrow}^\dagger
+\left(v_{m{\bf k}}^{(l)}\right)^* c_{m{\bf k}l\uparrow}
=S_l\hat{\gamma}_{em{\bf k}l\downarrow}^\dagger,
\nonumber 
\end{eqnarray}
where $S_l$ annihilates 
a coherent pair in the condensate of the $l$-th layer 
and $S_l^*$ creates one. It is not necessary to have 
the form of $b^\dagger_{m{\bf k}l}$. 
Inserting the inverse transformation to a 
pair-hopping process, we find that 
\begin{eqnarray}
\lefteqn{
c^\dagger_{m{\bf k}+{\bf p}l\uparrow}
c^\dagger_{m-{\bf k}-{\bf p}l\downarrow}
c_{m'{\bf k}l'\uparrow}
c_{m'-{\bf k}l'\downarrow}}
\nonumber \\
&=&
\left(
u_{m{\bf k}+{\bf p}}^{(l)}
\hat{\gamma}_{em{\bf k}+{\bf p}l\uparrow}^\dagger
+
\left(v_{m{\bf k}+{\bf p}}^{(l)}\right)^*
S_l^* \hat{\gamma}_{em{\bf k}+{\bf p}l\downarrow}
\right)
\left(
u_{m{\bf k}+{\bf p}}^{(l)}
\hat{\gamma}_{em{\bf k}+{\bf p}l\downarrow}^\dagger
-
v_{m{\bf k}+{\bf p}}^{(l)}
S_l^* \hat{\gamma}_{em{\bf k}+{\bf p}l\uparrow}
\right)
\nonumber \\
&\times&
\left(
u_{m'{\bf k}}^{(l')}
\hat{\gamma}_{em'{\bf k}l'\downarrow}
-
\left(v_{m'{\bf k}}^{(l')}\right)^*
S_{l'} \hat{\gamma}_{em'{\bf k}l'\uparrow}^\dagger
\right)
\left(
u_{m'{\bf k}}^{(l')}
\hat{\gamma}_{em'{\bf k}l'\uparrow}
+
v_{m'{\bf k}}^{(l')}
S_{l'} \hat{\gamma}_{em'{\bf k}l'\downarrow}^\dagger
\right)
\nonumber \\
&=&
\left(
(u_{m{\bf k}+{\bf p}}^{(l)})^2
\hat{\gamma}_{em{\bf k}+{\bf p}l\uparrow}^\dagger
\hat{\gamma}_{em{\bf k}+{\bf p}l\downarrow}^\dagger
+
u_{m{\bf k}+{\bf p}}^{(l)}
\left(v_{m{\bf k}+{\bf p}}^{(l)}\right)^*
S_l^*(
1
-\hat{n}_{em{\bf k}+{\bf p}l\downarrow}
)
\right.
\nonumber \\
&&
\left.
-
u_{m{\bf k}+{\bf p}}^{(l)}
v_{m{\bf k}+{\bf p}}^{(l)}
S_l^*
\hat{n}_{em{\bf k}+{\bf p}l\uparrow}
-
\left|v_{m{\bf k}+{\bf p}}^{(l)}\right|^2
(S_l^*)^2
\hat{\gamma}_{em{\bf k}l\uparrow}
\hat{\gamma}_{em{\bf k}l\downarrow}
\right)
\nonumber \\
&\times&
\left(
(u_{m'{\bf k}}^{(l')})^2
\hat{\gamma}_{em'{\bf k}l'\downarrow}
\hat{\gamma}_{em'{\bf k}l'\uparrow}
+
u_{m'{\bf k}}^{(l')}
v_{m'{\bf k}}^{(l')}
S_{l'}(
1-\hat{n}_{em'{\bf k}l'\downarrow})
-
u_{m'{\bf k}}^{(l')}
\left(v_{m'{\bf k}}^{(l')}\right)^*
S_{l'}\hat{n}_{em'{\bf k}l'\uparrow}
\right.\nonumber \\
&&
\left.
-
\left|v_{m'{\bf k}}^{(l')}\right|^2
S_{l'}^2 
\hat{\gamma}_{em'{\bf k}l'\uparrow}^\dagger
\hat{\gamma}_{em'{\bf k}l'\downarrow}^\dagger
\right).
\end{eqnarray}
Thus, we have a tunneling process from 
a Cooper pair in a condensate in a layer to a pair of 
quasi-electrons in the next layer by the contribution 
$(u_{m{\bf k}+{\bf p}}^{(l)})^2
u_{m'{\bf k}}^{(l')}
v_{m'{\bf k}}^{(l')}
\hat{\gamma}_{em{\bf k}+{\bf p}l\uparrow}^\dagger
\hat{\gamma}_{em{\bf k}+{\bf p}l\downarrow}^\dagger S_{l'}
+{\rm H.c.}$, which is found in the above expression.  

In a correlated quasi-electron system of the $l$-th superconducting 
layer, the charge fluctuation mode in 
${\cal H}_{i,{\rm intra}}^2$ produces 
short-range repulsive terms for the quasi-electron state 
given by $\hat{\gamma}_{em{\bf k}l\uparrow}^\dagger$ and 
$\hat{\gamma}_{em{\bf k}l\downarrow}^\dagger$. 
Since pair hopping occurs at a local position, 
the quasi-electrons in a pair inevitably raise their energy. 
To determine the localized nature of the hopping pair, 
we can do a rough estimation of the relative distance between 
two quasi-electrons in real space. 
For simplicity, we omit the radial dependences of 
$u_{m'{\bf k}}^{(l')}$ and $v_{m'{\bf k}}^{(l')}$, 
keeping only the energy dependences. 
Since we perform integrations in the $k$ space, 
by considering a superconducting gap much smaller than 
the Kohn-Sham-band width of a few eV, 
we have the following simplified expression for the creation operator 
of a hopping pair in an $m$-th band at an $l$-th layer 
from the next $l'$-th layer: 
\begin{eqnarray}
\lefteqn{
\int \! \! \int d^2 k d^2 p \;
(u_{m{\bf k}+{\bf p}}^{(l)})^2
u_{m'{\bf k}}^{(l')}
v_{m'{\bf k}}^{(l')}
\hat{\gamma}_{em{\bf k}+{\bf p}l\uparrow}^\dagger
\hat{\gamma}_{em{\bf k}+{\bf p}l\downarrow}^\dagger}
\nonumber \\
&\simeq&
\int \! \! \int d^2 k d^2 p \;
\theta(\varepsilon_{m,{\bf k}+{\bf p}}-E_F)
\delta(\varepsilon_{m',{\bf k}}-E_F)
\hat{\gamma}_{em{\bf k}+{\bf p}l\uparrow}^\dagger
\hat{\gamma}_{em{\bf k}+{\bf p}l\downarrow}^\dagger
\nonumber \\
&\simeq&
\int \! \! \int d^2 k d^2 p \;
\theta(\varepsilon_{m,{\bf k}+{\bf p}}-E_F)
\delta(\varepsilon_{m',{\bf k}}-E_F)
c_{m{\bf k}+{\bf p}l\uparrow}^\dagger
c_{m-{\bf k}-{\bf p}l\downarrow}^\dagger
\end{eqnarray}
Here, $E_F$ is the Fermi energy of the Kohn-Sham band structure 
and we have omitted the $k_z$ dependence of 
Kohn-Sham energy owing to its two-dimensional nature. 
If we further consider a semimetallic band with 
the $m$-th conduction band, the above expression 
indicates that a contribution of doubly occupied Wannier states appears. 
In iron arsenide, 
this Wannier state should be in a localized $3d$ state at an iron site. 

The above-mentioned characteristic feature of the Hubbard-type correlated 
system is very important in considering layered 
superconductors with 3d local orbitals. 
The pair-hopping process from the condensate to a 
correlated quasi-particle state 
leads to the conclusion of a discretized energy spectrum 
as a function of the number of hopping processes at a time. 
In other words, depending on the number 
of quasi-particle pairs, $n_{\rm pair}$, and 
$m_{\rm pair}$ of pair holes 
in a layer, we obtain the energy contribution of 
$E(n_{\rm pair},m_{\rm pair})=U_en_{\rm pair}+U_hm_{\rm pair}$ 
with the effective parameters $U_e>0$ and $U_h>0$. 
A simple correspondence of $n_{\rm pair}$ 
in the physics of the Hubbard model is shown by 
the number of doubly occupied sites. 

Let's therefore consider the situation given in 
Fig.\,\ref{Fig-1}(b), where 
only one pair is left from one layer to another layer. 
The bosonic nature of the pair allows us to write down 
an effective Hamiltonian in a Heisenberg spin system. 
Since we have a hopping pair or a vacancy at a layer, 
we have at least three states for each layer. 
We can assign these states to $S_z=\pm 1$ and $S_z=0$ 
of an artificial spin state and 
introduce an effective spin Hamiltonian. 
This minimal case is an $S=1$ system, in which the $l$-th layer 
may have one of these three states: $S_z=0$ and $\pm 1$. 

Now, the pair-hopping process is described by 
the $xy$-term in the Heisenbserg model. 
Neighboring charges with different signs will lower 
the energy, but two neighboring pairs with the same sign 
will raise the energy. This contribution is 
described by the anti-ferromagnetic 
$z$-term in the Heisenbserg exchange interaction. 
The charge neutrality discussed above also leads on-site anisotropy, 
{\it i.e.} the $D$ term, to stabilize the $S_z=0$ state, 
which corresponds to a neutral layer. 
These contributions are described by an XXZ model with the $D$ term. 
If a semimetallic band structure is the starting limit, 
and if the charge imbalance is minimized, 
the effective local charge neutrality is expressed 
by local energy enhancement, which is symmetric with 
quasi-particle pairs and pair holes, then a simple 
$D$ term should appear. In general, 
the charge neutrality condition can be effectively expressed 
by the introduction of a $D$ term. 
Thus, the minimal model is a one-dimensional 
anisotropic $S=1$ Heisenberg antiferromagnetic spin chain. 
\begin{equation}
{\cal H}_{\rm AFHM}
=\sum_{l}\left[
J(S_{l}^xS_{l+1}^x+S_{l}^yS_{l+1}^y)+J_zS_{l}^zS_{l+1}^z
\right]
+D\sum_{l}(S_{l}^z)^2.
\end{equation}

We know that there are three gapped phases in this model: 
the N\`{e}el phase, large-D phase, and 
Haldane phase.\cite{Haldane1,Haldane2} 
The N\`{e}el phase corresponds to a pair-vacancy array 
in the present model. The state thus corresponds to 
a charge density wave state along the stacking direction. 
Thus, it is not relevant for the present consideration 
for the superconductor. 
The large-D phase corresponds to a Mott insulating phase, 
where the charge fluctuation effect is suppressed. 
In the limiting case in the large-D phase, 
a decoupled array of two-dimensional electron gas is realized 
owing to the suppression of the charge fluctuation. 
For a relevant contribution to stabilize a bulk superconductor, 
the Haldane phase is the necessary phase. 
In the Haldane phase, a valence-bond-solid state\cite{AKLT} is realized 
with a broken hidden string order parameter.\cite{String1,String2} 
This gapped state with a hidden extra order contributes 
to the stabilization of bulk superconductivity. 
The phase diagram of the antiferromagnetic $S=1$ XXZ chain 
with uniaxial single-ion anisotropy is extensively studied.\cite{Numerical-gap-estimation1,Numerical-gap-estimation2,Numerical-gap-estimation3,Numerical-gap-estimation4,Numerical-gap-estimation5} 

The Haldane phase possesses an excitation gap. 
The lowest excitation with a total effective spin $S^z_{\rm tot}=1$ 
corresponds to the creation of an extra Cooper pair in the bulk. 
In this state, the number of bosons actually increases by one. 
When the effective model has $J=J_z=1$ with $D=0$, 
the gap becomes $\Delta \simeq 0.410479J$. 
There are continuous series of trials for determining Haldane gap. 
Recently, Ueda {\it et al.} have provided an estimation of 
lower and upper bounds 
and concluded that the gap is in $[0.41047905, 0.41047931]$.\cite{Ueda} 
This study is performed by the combined use of the hyperbolic-deformation 
technique and sequence interval squeeze method. 
When interaction parameters are varied, the gap changes continuously 
in the Haldane phase. 

Determining the value of the gap may be a simple test for checking 
the consistency of the present theory. 
If the gap is typically comparable to the transition temperature, 
we should roughly observe the pair hopping with $J\sim k_BT_c/0.410$. 
If $T_c$ is approximately 100\,K, $J$ should be approximately 24\,meV. 
This value is within reasonable range for the two-body 
effective interaction, which is derived as a charge fluctuation. 
On the other hand, $J_z$ and $D$ for 
diagonal elements of the effective model can have larger 
values without the electrostatic breakdown of vacuum or the insulating 
barrier layer between 
superconducting layers, {\it e.g.}, Fe$_2$As$_2$. 

Here, note that an $S_z=1$ state corresponds to 
any state with an extra pair in a layer, 
which may have multiple-colored states distinguished 
from each other. 
We only need to count the number of pairs that comes in 
or leaves a layer. 
If we consider states with many pairs coming in 
(or many vacancies leaving) a layer, 
we may utilize a higher-spin Heisenberg chain model. 
However, 
we know that Haldane gap decreases with increasing integer $S$. 
Thus, a high-temperature superconductor 
should be searched in a layered material, 
which can be mapped to an $S=1$ model. 
One might find that the above discussion is easily applied 
to triplet superconductors in layered materials, 
when inter-triplet-pair Coulomb fluctuation is taken into account. 

\section{Density Functional Variational Theory}
Recently, the author has proposed 
a theory of the model space in the multireference density functional 
theory.\cite{Kusakabe-model-space,Kusakabe-JPSJ,Kusakabe-Coulomb} 
In this formulation, we use a new variational principle 
for the electron models defined by the density functional theory. 
A version of the density functional variational theory is 
given by the nequality 
\begin{equation}
E_0\le
\min_{X_i,\varepsilon_i,g_i}\left\{
\min_{\Psi}
\bar{G}_{X_i,\varepsilon_i,g_i}[\Psi]
+
\Delta \bar{E}_{X_i,\varepsilon_i,g_i}[\Psi]
\right\},
\label{Variational}
\end{equation}
where $E_0$ is the ground-state energy of the electron system, and 
$\Psi$ inserted in $\Delta \bar{E}_{X_i,\varepsilon_i,g_i}[\Psi]$ 
is the minimizing $\Psi$ of the functional 
$\bar{G}_{X_i,\varepsilon_i,g_i}[\Psi]$.
The energy functional determining 
the model is given as  
\begin{eqnarray}
\bar{G}_{X_i,\varepsilon_i,g_i}[\Psi]
&=&
\langle \Psi | \hat{T}+\hat{V}_{X_i}|\Psi \rangle 
+\frac{e^2}{2}
\int d^3r d^3r' \frac{n_\Psi({\bf r})n_\Psi({\bf r}')}{|{\bf r}-{\bf r}'|}
\nonumber \\
&&+E_{\varepsilon_i}[n_\Psi]
+E_{g_i}[\Psi]+\int d^3r v_{ext}({\bf r})n_\Psi({\bf r}). 
\end{eqnarray}
Here, 
$E_{\varepsilon_i}[n]$ is an LDA energy functional, which may be
a GGA energy functional, and 
$E_{g_i}[\Psi]$ is a nonlocal correction parameter used in a standard 
DFT model. $E_{g_i}[\Psi]$ may be written by 
projection operators using a separable pseudo potential technique, 
an ultra-soft pseudo-potential technique, and 
a projector augmented wave technique.
We introduced an energy difference functional as 
\begin{eqnarray}
\Delta \bar{E}_{X_i,\varepsilon_i,g_i}[\Psi]
&=&
\frac{e^2}{2} \int d^3r d^3r' 
\frac{1}{|{\bf r}-{\bf r}'|}
\langle \Psi | 
:(\hat{n}({\bf r})-n_\Psi({\bf r}))(\hat{n}({\bf r}')-n_\Psi({\bf r}')):
|\Psi\rangle
\nonumber \\
&-&
E_{\varepsilon_i}[n_\Psi]
-E_{g_i}[\Psi]
-\langle \Psi| \hat{V}_{X_i}| \Psi \rangle .
\label{Delta_E}
\end{eqnarray}
The proof of eq.\,(\ref{Variational}) is easily realized by noting 
the next inequality 
\begin{eqnarray}
\Delta E_{X_i,\varepsilon_i,g_i}[\Psi]
&=&
F[n_\Psi]
-\frac{e^2}{2}
\int d^3r d^3r' \frac{n_\Psi({\bf r})n_\Psi({\bf r}')}{|{\bf r}-{\bf r}'|}
\nonumber \\
&-&
E_{\varepsilon_i}[n_\Psi]
-\min_{\Psi'\rightarrow n_\Psi}\left\{
\langle \Psi' | \hat{T}+\hat{V}_{red}^{X_i}| \Psi' \rangle 
+E_{g_i}[\Psi']\right\}
\nonumber \\
&=&
\frac{e^2}{2}\int d\lambda \int d^3r d^3r' 
\frac{1}{|{\bf r}-{\bf r}'|}
\langle \Psi_{n_\Psi}^\lambda | 
:(\hat{n}({\bf r})-n({\bf r}))(\hat{n}({\bf r}')-n({\bf r}')):
|\Psi_{n_\Psi}^\lambda\rangle
\nonumber \\
&-&
E_{\varepsilon_i}[n_\Psi]
+\min_{\Psi'\rightarrow n_\Psi}
\langle \Psi' | \hat{T} | \Psi' \rangle 
-\min_{\Psi'\rightarrow n_\Psi}\left\{
\langle \Psi' | \hat{T}+\hat{V}_{red}^{X_i}| \Psi' \rangle 
+E_{g_i}[\Psi']\right\} 
\nonumber \\
&=&
\min_{\Psi'\rightarrow n_\Psi}
\langle \Psi' | \hat{T} + \hat{V}_{\rm ee}| \Psi' \rangle 
-\frac{e^2}{2}
\int d^3r d^3r' \frac{n_\Psi({\bf r})n_\Psi({\bf r}')}{|{\bf r}-{\bf r}'|}
\nonumber \\
&-&
E_{\varepsilon_i}[n_\Psi]
-
\langle \Psi | \hat{T}+\hat{V}_{red}^{X_i}| \Psi \rangle 
-E_{g_i}[\Psi]
\nonumber \\
&\le&
\langle \Psi | \hat{T} + \hat{V}_{\rm ee}| \Psi \rangle 
-\frac{e^2}{2}
\int d^3r d^3r' \frac{n_\Psi({\bf r})n_\Psi({\bf r}')}{|{\bf r}-{\bf r}'|}
\nonumber \\
&-&
E_{\varepsilon_i}[n_\Psi]
-E_{g_i}[\Psi']
-\langle \Psi | \hat{T}+\hat{V}_{red}^{X_i}| \Psi \rangle 
\nonumber \\
&=&
\langle \Psi | \hat{V}_{\rm ee}| \Psi \rangle 
-\frac{e^2}{2}
\int d^3r d^3r' \frac{n_\Psi({\bf r})n_\Psi({\bf r}')}{|{\bf r}-{\bf r}'|}
\nonumber \\
&-&
E_{\varepsilon_i}[n_\Psi]-E_{g_i}[\Psi']
-\langle \Psi | \hat{V}_{red}^{X_i}| \Psi \rangle 
\nonumber \\
&=& \Delta \bar{E}_{X_i,\varepsilon_i,g_i}[\Psi]. 
\end{eqnarray}
Here, we used 
\begin{equation}
\min_{\Psi'\rightarrow n_\Psi}\left\{
\langle \Psi' | \hat{T}+\hat{V}_{red}^{X_i}| \Psi' \rangle 
+E_{g_i}[\Psi']
\right\} 
=
\langle \Psi | \hat{T}+\hat{V}_{red}^{X_i}| \Psi \rangle 
+E_{g_i}[\Psi],
\end{equation}
since the minimizing $|\Psi'\rangle$ of the above expression 
for the charge density $n_\Psi({\bf r})$ is 
obtained by minimizing a functional, 
\begin{eqnarray}
\hat{G}_{n_\Psi}[\Psi']
&=&
\langle \Psi' | \hat{T}+\hat{V}_{X_i}|\Psi' \rangle 
+\frac{e^2}{2}
\int d^3r d^3r' \frac{n_\Psi({\bf r})n_\Psi({\bf r}')}{|{\bf r}-{\bf r}'|}
\nonumber \\
&&+E_{\varepsilon_i}[n_\Psi]
+E_{g_i}[\Psi']+\int d^3r v_{ext}({\bf r})n_\Psi({\bf r}). 
\end{eqnarray}
which is given by $|\Psi\rangle$. 
We again find importance of the self-consistency in 
the minimizing process of 
$\bar{G}_{X_i,\varepsilon_i,g_i}[\Psi]$.

Thus,
\begin{eqnarray}
E_0
&\le&
\min_{X_i,\varepsilon_i,g_i}\left\{
\min_{\Psi}
\bar{G}_{X_i,\varepsilon_i,g_i}[\Psi]
+
\Delta {E}_{X_i,\varepsilon_i,g_i}[\Psi]
\right\}
\nonumber \\
&\le&
\min_{X_i,\varepsilon_i,g_i}\left\{
\min_{\Psi}
\bar{G}_{X_i,\varepsilon_i,g_i}[\Psi]
+
\Delta \bar{E}_{X_i,\varepsilon_i,g_i}[\Psi]
\right\}.
\nonumber 
\end{eqnarray}

Thus, we can start from a known 
LDA functional to construct a variational model of the electron system. 
The fluctuation term $\hat{V}_{X_i}$ is formed 
by static two-body correlation functions, which yield 
two-body effective interactions in the MR-DFT model. 
Thus, we have a firm ground for a beyond-LDA approach, 
considering that relevant fluctuation modes 
are inserted in the variational model. 
We may utilize GGA in place of LDA under the assumption that 
the exchange-correlation potential is always given in the simulation 
process. 

In the density functional variational theory (DFVT), 
a simulation for determining the fluctuation term 
$\langle \Psi | \hat{V}_{X_i} | \Psi \rangle$ 
is given, when a differentiable $E_{\varepsilon_i}[n]$ and/or 
an explicit $E_{g_i}[\Psi]$ are prepared. 
Then, we always have a defined one-body part of the effective Hamiltonian 
in a self-consistent determination process of 
the self-consistent solution of the model. 
Thus, an LDA or GGA solution can be used to construct 
a correlated electron model of superconductivity. 
The model has effective two-body interaction terms. 
According to the model Hamiltonian, 
we can apply any appropriate solver for an effective many-body problem. 
Self-consistency is 
imposed by calculating the charge density, 
which redefines the exchange-correlation potential. 
Two-body processes and their interaction 
parameters are determined, so that they reduce the variational 
energy of the Coulomb system. The variational energy is given by 
evaluating all the terms in eqs.\,(\ref{Variational}) and (\ref{Delta_E}). 
At present, realistic determination processes are 
computationally demanding, 
but a preliminary simulation of Sr$_2$CuO$_3$\cite{Sogo} indicates that 
an optimization process indeed works for determining 
an effective interaction parameter. 
In this one-dimensional copper oxide with the $d^9$ configuration 
at each Cu, the on-site Hubbard interaction $U>0$ is determined 
by searching the minimum variational energy. 

\section{Conclusions}
We proposed a pair-hopping mechanism expected in 
a layered superconductor on the basis of MR-DFT. 
The definition of the effective model for a superconductor 
is given from the first-principles method. 
Determination techniques of the model is given by DFVT. 
The derived effective pair-hopping model suggests 
the realization of a valence-bond-solid state of 
an effective $S=1$ Heisenberg anti-ferromagnetic spin chain. 
A bulk superconducting phase becomes gapped in the charge 
fluctuation mode. 

Sufficient conditions for the pair-hopping mechanism are summarized 
as follows: 
(i) There is a two-body pair hopping process between 
layers, which may be direct or indirect. 
(ii) Interlayer single-particle tunneling is negligible 
compared with the two-body pair hopping between layers. 
(iii) A correlation effect in a layer suppresses 
multiple-pair hopping at the same time, keeping 
the local charge neutrality and allowing minimum charge fluctuation 
in a uniform bulk superconducting state. 
These conditions allow us to have a high-temperature superconductor. 
The realization of the mechanism in iron arsenides is expected 
because we obtain (1) experimental observation of high-temperature 
superconductivity, (2) the two-dimensional electronic state 
given by GGA, and (3) the present formulation of the pair-hopping 
mechanism in MR-DFT. 
DFVT suggests that, if competing diagonal orders in 
magnetic and non-magnetic channels are not comparable 
in variational energy, 
the superconducting state is selected. 
For the off-diagonal superconducting order, 
there is energy reduction in Coulomb energy, 
because the pair-hopping mechanism selects 
the unique ground state with the Haldane gap. 

The importance of the interlayer pair tunneling process 
has often been stressed for cuprate 
high-temperature superconductors.\cite{Chakravarty1,Anderson,Chakravarty2}
In our DFVT, we can also derive a strategy 
for enhancing the stability of the superconductivity 
using our microscopic effective model. 
The Coulomb-originated pair hopping is derived via 
the charge fluctuation modes. 
One possible form is given by eq.\,(\ref{X_i-term}). 
Here, we need to consider the sign of the superconducting 
order parameter. 
We have two contributions, namely, 
a $\hat{Y}$-process and a $\hat{Z}$-process, due to 
the appearance of two operators. 
The important point is that the sign of these processes 
are different. 
Depending on the two-dimensional superconducting order parameter, 
one of them can be effective for direct pair hopping. 
As for the super pair tunneling mechanism, 
the combination of the $\hat{Y}$-process and $\hat{Z}$-process 
may appear in the higher-order pair hopping process from one layer 
to another layer. 

\begin{figure}[tb]
\begin{center}
\includegraphics[width=8.0cm]{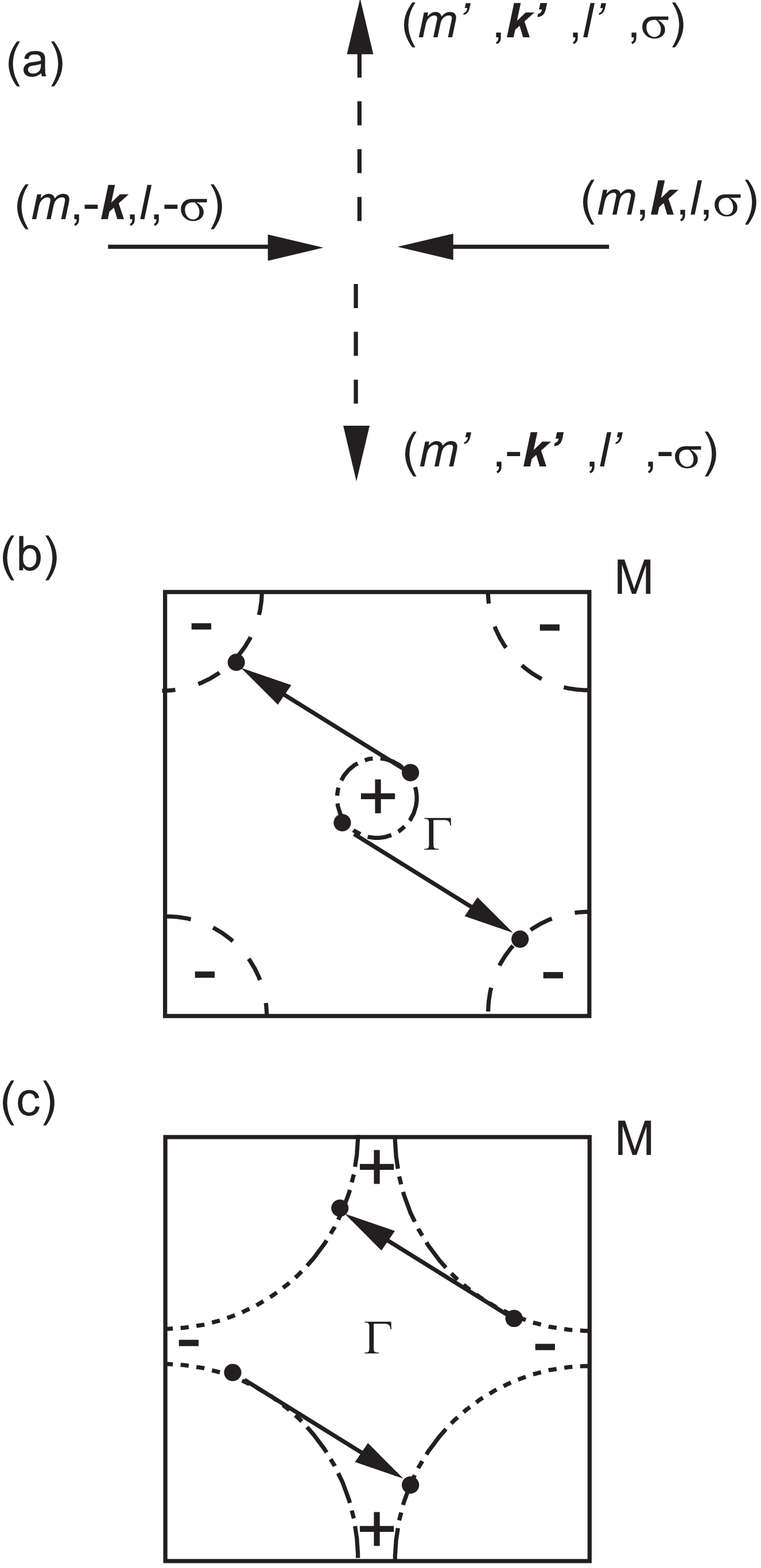}
\end{center}
\caption{
(a) 
Effective interaction process for the pair hopping from 
one layer to another layer. 
In the first $l$-th layer, a Cooper pair experiencing a positive pair potential 
is scattered into the next $l'$-th layer by a $\hat{Y}$-process, 
where the final state may be affected by a negative pair potential. 
In the scattering process, an electron with 
$(m,{\bf k},l,\sigma)$ of a pair 
is scattered into $(m',{\bf k}',l',\sigma)$ 
with a momentum shift 
and another electron $(m,-{\bf k},l,-\sigma)$
of the pair obtains an inverse momentum shift 
to be $(m',-{\bf k}',l',-\sigma)$. 
(b) 
Schematic viewgraph of the superconducting order parameter 
in the two-dimensional first Brillouin-zone of 
a model two-band superconducting state. 
If the Fermi surfaces are separated into a center one and 
another one around the $M$ point, an extended $s$-wave state 
may be created. The interlayer pair hopping 
from the $l$-th center Fermi surface to 
the $l'$-th Fermi surface around $M$ may occur. 
(c) 
A schematic viewgraph representing the sign of the order parameter 
for a model superconducting state with a $d$-wave state. 
}
\label{Fig-2}
\end{figure}

In Fig.\,\ref{Fig-2}(a), 
we show a possible scattering process from one layer to 
another layer. By using a $\hat{Y}$-process, 
two branches of the gap function, {\it i.e.}, an $m$-th band 
with a positive sign and another $m'$-th band with a negative sign 
may be stabilized. 
In a $\hat{Z}$-process, the final pair potential does not 
need to change the sign from the initial one. 
The momentum conservation at a scattering center holds. 
According to this rule, we can find out relevant 
scattering processes depending on the geometry of the 
Fermi surfaces and the sign of the local order parameter 
$\bar{\Delta}_{jj'l}$. 
Two examples shown in Fig.\,\ref{Fig-2} 
are for (b) an extended $s$-wave state modeling iron arsenides 
and (c) a $d$-wave state for cuprates. 
The strength of these processes is dependent 
on the orbitals $\phi_{jl}$ and effective coupling strength, 
and thus on the material structure considered. 
Furthermore, we need a strong superconducting fluctuation 
in a two-dimensional layer. For the enhancement of the intralayer 
fluctuation, we may rely on our knowledge of the spin-fluctuation mechanism 
of high-temperature superconductivity.\cite{Monthoux,Arita} 
Another known fact in the literature supporting the present pair-hopping 
mechanism is the electronic band structure calculations 
for optimally doped LaFeAsO,\cite{Mazin2} which has
a higher $T_c$ than NaFeAs. 
The two-dimensionality of Fermi surfaces is confirmed, 
when the experimentally observed lattice structure is assumed 
in the simulation or when hole doping is assumed in a theoretically 
determined lattice structure. 
A perfect two-dimensionality of the LDA or GGA band structures 
means a strong suppression of 
interlayer single-particle hopping processes as well as 
the appearance of correlation effects in each layer. 
The indirect pair tunneling mechanism further suggests 
the co-existence of magnetic order in an independent part of the 
layered superconducting system. If the magnetic structure only 
acts as a medium supporting the pair hopping processes, 
the gap formation in a stack of local superconducting two-dimensional 
electron systems is expected. 

A simple comment on another attempt\cite{Yoshida} 
to elucidate a superconductivity is given for pedagogical reason. 
The present pair hopping mechanism is allowed only for 
the interlayer pair scattering due to the Coulomb repulsion 
between electrons. If there is an effective attractive 
interaction at an attractive center between layers, 
the transition temperature vanishes, since the pair hopping is blocked. 
The existence of a on-site static attractive interaction 
coming from the Coulomb repulsion 
has been already disproved.\cite{Kusakabe-at} 

Finally, three possible comments on real superconductors are given as follows. 
The origin of inter-layer scattering processes relevant 
to bulk superconductivity may be the electron-phonon interaction 
as well as the intra-layer effective attraction. 
In the case of MgB$_2$,\cite{MgB2} 
the pair-hopping mechanism can give some amount of 
stabilization through the appearance of a hidden order parameter. 
Since we need to specify details of scattering processes 
that stabilize the Coulombic electron system, 
further theoretical investigation might be necessary. 

For the realization of the effective $S=1$ Heisenberg model, 
we conjectured that a semimetallic band structure is favorable. 
Pseudo-electron-hole symmetry corresponds 
to the XXZ Heisenberg chain with the $D$ term. 
To have the most stable Haldane gap, we also need to perform 
numerical simulation of a generalized Heisenberg chain model. 
Although a detailed discussion on 
the stability of the multiple-ordered state proposed on the basis of 
the present pair-hopping mechanism is required to determine 
the best condition by first-principles simulation,  
we expect to obtain a reliable estimator of $T_c$ in the near future, 
since the numerical accuracy for the determination of 
the Haldane gap is now going beyond single precision.\cite{Ueda} 

In several cuprate superconductors, 
the realization of microscopic Josephson junction arrays 
is known to be related to the Josephson 
plasma.\cite{Ozyuzer,Kadowaki,Lin-Hu,Hu-Lin}
For the microscopic analysis of this phenomenon, 
the determination of the sign and structure of 
the superconducting order parameter is important. 
A simulation based on DFVT is expected to solve 
this problem, too. 

\section*{Acknowledgment}
The author thanks A. Nakanishi, who showed his simulation data 
prior to the publication and allowed for referencing information. 
He is also grateful for stimulating discussion with 
Mr. H. Ueda and Prof. I. Maruyama. 
The present work is partly supported by Grant-in-Aid 
for Scientific Research from the Ministry of Education, Culture, 
Sports, Science and Technology of Japan (Grants No. 19051016), 
the Global COE Program (Core Research and Engineering of Advanced 
Material-Interdisciplinary Education Center for Materials Science), 
MEXT, Japan, 
and Grand Challenges in next-generation integrated nanoscience. 

\end{document}